\documentclass[showpacs,preprintnumbers,amsmath,amssymb,floatfix]{revtex4}
\usepackage{graphicx}% Include figure files
\usepackage{dcolumn}% Align table columns on decimal point
\usepackage{bm}% bold math

\begin{document}
%\preprint{APS/123-QED}

\title{Probing nuclear compressibility via fragmentation \\ in Au+Au reactions at 35 AMeV}
% Force line breaks with \\

\author{Yogesh K. Vermani}
\email{yugs80@gmail.com}
%\altaffiliation [Also at]{Physics Department, XYZ University.}%Lines break automatically or can be forced with \\
\author{Rajiv Chugh}
 \affiliation{Department of Physics, Panjab
University, Chandigarh -160 014, India}
\author{Aman D. Sood}%
 \affiliation{SUBATECH - Ecole des Mines de Nantes \\
4, rue Alfred Kastler, F-44072 Nantes, Cedex 03, France}

%\author{Charlie Author}
%\homepage{http://www.Second.institution.edu/~Charlie.Author}
%\affiliation{
%Second institution and/or address\\
%This line break forced% with \\

\date{\today}% It is always \today, today,

\begin{abstract}
The molecular dynamics study of fragmentation in peripheral
$^{197}$Au +$^{197}$Au collisions at 35 MeV/nucleon is presented
to probe the nuclear matter compressibility in low density regime.
The yields of different fragment species, rapidity spectra, and
multiplicities of charged particles with charge $3\leq Z \leq 80$
are analyzed at different peripheral geometries employing a soft
and a hard equations of state. Fragment productions is found to be
quite insensitive towards the choice of nucleon-nucleon cross
sections allowing us to constrain nuclear matter compressibility.
Comparison of calculated charged particle multiplicities with the
experimental data indicates preference for the \emph{soft} nature
of nuclear matter.
\end{abstract}


\maketitle

\section{INTRODUCTION}

During the last two decades, numerical simulations of medium and
high energy heavy-ion (HI) reactions have provided a unique
opportunity to explore the nuclear matter at the extreme
conditions of density and temperature
\cite{leh93,khoa,sk99,sod99,aich87}. The compression of nuclear
matter can be judged via equation of state (EoS) which is also a
main input for any theoretical model along with nucleon-nucleon
(\emph{n-n}) cross section \cite{sod99, aich87, peil, breng, part}. Various attempts have been made to find the observables
which are sensitive to the nuclear EoS. In the past, comparison of
theoretical predictions with the experimental results has been
used to extract the nuclear equation of state. One of the earlier
attempts for nuclear EoS (or, incompressibility) was via giant
monopole resonance (GMR) studies \cite{young, itoh}. The
scattering of $\alpha$ particles off the nucleus induces volume
oscillations with L=0, which can be used to determine the
incompressibility `$\kappa$' of that nucleus. These results
generally yield incompressibility in the range $\kappa\sim$
250-270 MeV indicating the matter to be softer. A recent GMR study
in the $^{208}$Pb and $^{90}$Zn nuclei showed that the softening
of nuclear matter is needed to explain the collective modes with
different neutron-to-proton ratios \cite{todd}. Another study on
the fusion reported linear momentum transfer to be sensitive to
both the EoS and \emph{n-n} cross section \cite{cib}. Within
\emph{quantum molecular dynamics} (QMD) model, an
incompressibility of $\kappa$=200 MeV (\emph{i.e.}, soft EoS) was
reported to reproduce the experimental data on energy transfer in
a compound nucleus formation \cite{cib}.

\par Collective flow observed in HI collisions is another observable which is found to be
sensitive towards the stiffness of nuclear EoS \cite{peil, part,
mages}. The collective transverse in-plane flow and balance energy
(the energy at which flow becomes zero) have been studied
extensively over the past two decades so as to constrain the EoS,
but still the uncertainties are very large. For example, a stiff
EoS with $\kappa$=380 MeV reproduces the transverse flow data
equally well as obtained with soft momentum dependent EoS with
$\kappa$=210 MeV \cite{aich87,pan}. Similarly, comparison
of transport model calculations with data of EOS Collaboration for
the energy dependence of collective flow favored neither the
`soft' nor the `hard' equation of state \cite{part}. In recent
comparison of elliptic flow data with microscopic transport model
calculations of the Refs. \cite{ell, ell1}, no consistent
agreement with the data could be obtained \cite{andro} for the two
different models of Refs. \cite{ell} \& \cite{ell1}. Therefore, it
is clear from the above review that an appropriate choice of the
nuclear equation of state is still far from settlement. The task
of deriving quantitative information about the EoS requires
detailed comparison of theoretical calculations assuming different
equations of state with experimental data.

At lower beam energies, Pauli blocking of final states gets more
pronounced. As a result, mean-field effects and long range Coulomb
force govern the reaction dynamics. If one goes to still lower
energy regime, well-known phenomena like complete fusion,
incomplete fusion, fission, cluster emission \emph{etc.}, can be
seen \cite{rkg,rkg1}. At incident energies above 20 AMeV, phenomena
like production of intermediate mass fragments (IMFs),
projectile-like and target-like fragments (PLFs and TLFs) dominate
the exit channel. The phenomenon of multifragment-emission in low
energy domain is, however, least exploited to infer the nuclear
EoS. Naturally the study of fragment-emission in low energy domain
may be of importance to constrain the nuclear incompressibility,
where the role of different \emph{n-n} cross sections is expected
to be minimal. To explore the possibility of achieving information
on the nuclear EoS, we plan here to simulate the peripheral
reactions of $^{197}$Au +$^{197}$Au at $E_{lab}$=35 AMeV and at
different peripheral geometries where accurate data has been
measured on Multics-Miniball set-up \cite{expt}. To this end, we
performed detailed calculations within the \emph{quantum molecular
dynamics} (QMD) model \cite{hart, aich91} which is described in
detail in section~\ref{model} along with \emph{simulated annealing
clusterization algorithm} \cite{saca,euro}. Section~\ref{results}
presents results of our numerical calculations and comparison with
available experimental data, which are finally summarized in
section~\ref{summary}.

\section{\label{model}The Model}

\subsection{Quantum Molecular Dynamics (QMD) Model}

The \emph{quantum molecular dynamics} model is a \emph{n}-body
transport theory that incorporates the quantum features of Pauli
blocking and stochastic \emph{n-n} scattering. Each nucleon in the
colliding system is represented by a Gaussian wave packet as:
\begin{equation}
{\psi}_i({\bf r},{\bf p}_i(t),{\bf r}_i(t))=\frac{1}{(2\pi
L)^{3/4}} exp \left[ \frac{i}{\hbar} {\bf p}_i(t)\cdot {\bf
r}-\frac{({\bf r}-{\bf r}_i(t))^2}{4L} \right]. \label{s1}
\end{equation}
Mean position $r_{i}(t)$ and mean momentum $p_{i}(t)$ are the two
time dependent parameters. The Gaussian width $\sqrt{L}$ is fixed
with a value of 1.8 fm and is same for all nucleons. This value of
$\sqrt{L}$ corresponds to a root-mean-square radius of each
nucleon. The centers of these Gaussian wave packets in ${\cal
R}_3$ and ${\cal P}_3$ spaces follow the trajectories according to
the classical equations of motion:
\begin{equation}
\dot{{\bf p}_i}=-\frac{\partial \langle {\cal H} \rangle}{\partial
{\bf r}_i}, ~ \dot{{\bf r}_i}=\frac{\partial \langle {\cal H}
\rangle}{\partial {{\bf p}_i}}. \label{euler}
\end{equation}
The Hamiltonian ${\cal H}$ appearing in Eq. (\ref{euler}) has
contribution from the local Skyrme-type, Yukawa and effective
Coulomb interactions \cite{hart}:
\begin{subequations}
\label{pot:whole}
\begin{eqnarray}
V_{ij}^{loc}&=& t_{1}\delta({\bf r}_i-{\bf r}_j) +
t_{2}\delta({\bf r}_i-{\bf r}_j)\delta({\bf r}_i-{\bf r}_k)  \\
V_{ij}^{Yuk}&=&t_3 \frac{exp \{ -| {\bf r}_i-{\bf r}_j|\}/\mu}{|{\bf r}_i-{\bf r}_j|/\mu} \\
V_{ij}^{Coul}&=& \frac {{Z_i}\cdot{Z_j}~e^2}{|{\bf r}_i -{\bf r}_j|}.
\end{eqnarray}
\end{subequations}
Here $Z_{i}$, $Z_{j}$ are the effective charge of baryons {\it i} and
{\it j}. The long-range Yukawa force is necessary to improve the surface
properties of the interaction. The parameters $\mu, t_{1}, t_{2},
t_{3}$ appearing in Eqs. (\ref{pot:whole}) are given in Ref. \cite{aich91}.
These parameters are adjusted and fitted so as to achieve the
correct binding energy and root mean square values of the radius
of the nucleus \cite{aich91}. In QMD model, one neglects the isospin dependence of nucleon-nucleon interaction.
All nucleons in a nucleus are assigned the effective
charge $Z=\frac{Z_{T}+Z_{P}}{A_{T}+A_{P}}$ \cite{hart}. It is worth
mentioning that isospin dependent flavor of QMD model (\emph{i.e.} IQMD)
has also been used in literature \cite{hart,saks}. This microscopic transport code explicitly takes into account
the differences of neutron and proton potentials and cross sections. The Skyrme part of the interaction used in QMD model
has the generalized form:
\begin{equation}
V^{loc}_{ij}=\frac{\alpha}{2}\left(\frac{\rho_{ij}}
{{\rho}_{\circ}} \right)+ \frac{\beta}{\gamma
+1}\left(\frac{\rho_{ij}}{{\rho}_{\circ}} \right)^\gamma.
\label{Sk}
\end{equation}
The interaction density in Eq. (\ref{Sk}) is defined as:
\begin{equation}
\rho_{ij}=\frac{1}{(4\pi L)^{3/4}}\cdot e^{-({\bf r}_i-{\bf
r}_j)^2/4L}.
\end{equation}
The parameters $\alpha$ and $\beta$ are adjusted to reproduce the
infinite nuclear matter binding energy (E/A = -16 MeV) at
saturation nuclear matter density $\rho_{\circ}$. The third
parameter $\gamma$ can be varied independently to account for
different nuclear incompressibilities $\kappa$ (\emph{i.e.}
different equations of state). Two different parameterizations are
used: a soft EoS with incompressibility $\kappa$=200 MeV, and a
hard EoS with $\kappa$=380 MeV. The parameters $\alpha,~\beta,~and~\gamma$ employed in the
QMD model serve as an input for repulsive potential with high compressibility (\emph{i.e.} hard EoS),
and less repulsive potential (\emph{i.e.} soft EoS).
The standard parameters corresponding to these two equations
of state are listed in Table~\ref{param}.
\begin{table}[!h]
\centering \caption{\label{param} The Skyrme parameters for `soft' and `hard' interactions
used in the QMD model.} \vspace
{0.32cm}
%\begin{ruledtabular}
\begin{tabular}{lllll}
\hline \hline
$EoS$ & $\alpha$(MeV) & $\beta$(MeV) & $\gamma$ & $\kappa$(MeV) \\
\hline

Soft & -356 & 303 & 7/6 & 200 \\
Hard & -124 & 70.5 & 2 & 380 \\
\hline \hline
\end{tabular}
%\end{ruledtabular}
\end{table}

The influence of different \emph{n-n} scattering cross sections will be determined by
employing a set of different cross sections varying from
energy-dependent cross section \cite{cug81} to constant and
isotropic cross sections with magnitudes of 40 and 55 mb. It is
worth mentioning that in recent times, even relativistic version
has also been analyzed \cite{leh93}. As noted in Ref.
\cite{leh93}, this has no effect on present findings. A hard EoS
with energy dependent cross section is labeled as $Hard^{Cg}$.
Incorporation of isotropic and constant cross sections of 40 and
55 mb strengths have been labeled as $Hard^{40}$ and $Hard^{55}$,
respectively. Similarly, for the soft equation of state, we have
$Soft^{Cg}$, $Soft^{40}$ and $Soft^{55}$, respectively. Since QMD
model follows the time evolution of nucleons only, one has to
employ secondary clusterization algorithms to identify fragments'
structure. In the present paper, \emph{simulated annealing
clusterization algorithm} (SACA) has been used to identify final
fragment structure. This method is reported to explain the ALADiN
data on spectator fragmentation quite nicely at relativistic
bombarding energies \cite{euro}.

\subsection{\label{esaca}The SACA Formalism}

This clusterization procedure allows early identification of
fragments before these are well separated in coordinate space. In
the SACA method, fragments are constructed based on the energy
correlations. It works on the principle of energy minimization of
fragmenting system. Standard \emph{minimum spanning tree} (MST) procedure \cite{aich91}
is employed to obtain pre-clusters. This procedure assumes that nucleon pairs separated by distance $\mid{\bf
r}_{i}-{\bf r}_{j}\mid \leq$ 4 fm, belong to the same fragment \cite{aich91, peil}.
Such preliminary cluster configuration is determined at every time step.
Thus one can address the time evolution of mass, charge, position and momentum of each fragment or single nucleon.
The pre-clusters obtained with the MST method \cite{aich91, jai} are subjected to a binding energy check \cite{saca}:
\begin{eqnarray}
\zeta_{i}=\frac{1}{N_{f}}\sum_{\alpha=1}^{N_{f}}
\left[\sqrt{\left(\textbf{p}_{\alpha}-\textbf{P}_{N_{f}^{c.m.}}
\right)^{2}+m_{\alpha}^{2}}-m_{\alpha} \right.
+\left.\frac{1}{2}\sum_{\beta \neq \alpha}^{N_{f}}V_{\alpha \beta}
\left(\textbf{r}_{\alpha},\textbf{r}_{\beta}\right)\right]<
-E_{bind}, \label{be}
\end{eqnarray}
with $E_{bind}$ = 4.0 MeV if $N_{f}\geq3$ and $E_{bind} = 0$
otherwise. In Eq. (\ref{be}), $N_{f}$ is the number of nucleons in
a fragment and $\textbf{P}_{N_{f}}^{c.m.}$ is the center-of-mass
momentum of the fragment. The requirement of a minimum binding
energy excludes the loosely bound fragments which will decay at
later stage. To look for the most bound configuration (MBC), we
start from a random configuration which is chosen by dividing
whole system into few fragments. The energy of each cluster is
calculated by summing over all the nucleons present in that
cluster using Eq. (\ref{be}).

Let the total energy of a configuration k be $E_{k}(=
{\sum_{i}}N_{f}\zeta_{i})$, where $N_{f}$ is the number of
nucleons in a fragment and $\zeta_{i}$ is the energy per nucleon
of that fragment. Suppose a new configuration $k^{'}$ (which is
obtained by (a)transferring a nucleon from randomly chosen
fragment to another fragment or by (b) setting a nucleon free, or
by (c) absorbing a free nucleon into a fragment) has a total
energy $E_{{k}^{'}}$. If the difference between the old and new
configuration $\Delta E (= E_{{k}^{'}}-E_{k})$ is negative, the
new configuration is always accepted. If not, the new
configuration $k^{'}$ may nevertheless be accepted with a
probability of $exp (-\Delta E/\upsilon)$, where $\upsilon$ is
called the control parameter. This procedure is known as
Metropolis algorithm. The control parameter is decreased in small
steps. This algorithm will yield eventually the most bound
configuration (MBC). Since this combination of a Metropolis
algorithm with slowly decreasing control parameter $\upsilon$ is
known as \emph{simulated annealing}, so our approach is dubbed as
\emph{simulated annealing clusterization algorithm} (SACA). We
have used here an extended version of SACA, in which each cluster
is subjected to its true binding energy based upon modified
Bethe-Weizs\"{a}cker mass formula \cite{yugs}. It may be stated
that fragmentation analysis performed within extended version
yields the same results as with constant binding energy check of
-4 MeV/nucleon. It also justifies using extended approach to
analyze the fragmentation at such a low incident energy when
nuclei are still in Fermi energy domain. The constant binding energy criterion (of -4 MeV/nucleon) is chosen keeping in
mind the average binding energy of clusters. We have also analyzed the fragmentation pattern employing $E_{bind}$ based upon
experimental binding energies. Nearly no effect of this modification was found. For further details, we refer the reader to Refs. \cite{saca,euro,yugs, jai}.

\section{ \label{results} Results and Discussion}

Figure 1 shows the time evolution of $^{197}$Au +$^{197}$Au
reaction at 35 AMeV with reduced impact parameters
$b/b_{max}$=0.55 (left panel) and 0.85 (right panel). Top panel
depicts the time evolution of average nucleon density $\rho^{avg}$
for soft and hard EoS. One notices several interesting results:
(i) Maximal density is reached nearly at the same time at both
impact parameters, whereas saturated values is slightly more for
higher impact parameters, (ii) Choice of different \emph{n-n}
cross sections have insignificant influence on the results
obtained. This happens due to effective Pauli blocking at such a
low incident energy that prohibits \emph{n-n} collisions.

Stiffness of nuclear EoS, however significantly influences the
mean nucleon density $\rho^{avg}$ and other fragment observables
shown in subsequent windows. This difference is clearly visible in
the evolution of heaviest fragment $\langle A^{max} \rangle$,
multiplicities of free particles, light charged particles LCPs [$2
\leq A \leq 4$], and clusters with mass $A \geq 5$. The mean size
of heaviest fragment $\langle A^{max} \rangle$ attains minimum
around 100 fm/c, where stable fragment configuration can be
realized and compared with experimental results. With stiff EoS,
heavier $\langle A^{max} \rangle$ is registered. Interestingly,
multiplicity of free particles obtained also follows the same
trend as $\langle A^{max} \rangle$. This means that dissipation of
energy takes place mainly via emission of free-nucleons that cools
down the nuclear system in case of hard EoS. Consequently, lesser
yields of LCPs and fragments with mass $A\geq5$ are obtained with
a stiff EoS. On other hand, soft EoS favors emission of LCPs and
heavier fragments ($A\geq5$) from spectator zone, thereby
decreasing the size of $A^{max}$. The insensitivity of
fragmentation pattern towards choice of different \emph{n-n} cross
sections may be, therefore, useful to constrain the nuclear
compressibility in low density regime.

Next we study the rapidity spectra of free nucleons and
intermediate mass fragments in transverse and longitudinal
direction using a hard and a soft equations of state. The transfer
of excitation energy from the participant zone to spectator matter
has a direct bearing on the rapidity distribution of the
fragments. Figure 2 displays the spectrum of scaled transverse $y^{(x)}$  and longitudinal $y^{(z)}$ rapidity distribution
of free particles (top) and intermediate mass fragments IMFs [$5
\leq A \leq 65$] (bottom) for the collision of Au (35 AMeV)+ Au at
reduced impact parameter $b/b_{max}$=0.55. As expected, the
rapidity spectrum of free nucleons and IMFs is quite sensitive to
the nuclear EoS that brings out significant change in their
transverse expansion as well as stopping pattern. Using hard
interactions, a larger fraction of free nucleons are emitted into
transverse direction. IMFs aren't, however, dispersed much into
transverse directions and continue to move at target and
projectile velocities. Similar trends are visible for longitudinal
rapidity ($y_{z}$) distribution as well. Using a `stiff' EoS,
system seems to cool-off via abundant production of the free
nucleons from the midrapidity as well as from the spectator zone,
whereas a `soft' EoS contributes significantly towards the IMFs
emission at target and projectile rapidities. It means that system
propagating under the soft interactions is less equilibrated. As a
result, heavier fragments leave the participant zone quite early
and suffer less collisions. These findings suggest that fragment
emission from the decay of spectator component is quite sensitive
to the mean field and compressibility of participant matter.

Finally we calculate the multiplicity of charged particles with
$3\leq Z \leq 80$ using a hard and a soft equations of state for
Au(35 AMeV)+Au collisions at six peripheral geometries (see Fig.
3). The multiplicities calculated at 100 fm/c are subjected to
forward rapidity condition (y $>$ 0.5 y$_{beam}$) in the
center-of-mass frame to exclude events from midrapidity and
quasitarget decay. Also shown in the figure is integrated
multiplicities of charged particles with charge $3\leq Z \leq 80$
(\emph{i.e.} $\int_{3}^{80}N(Z)dZ$) obtained on Multics-Miniball
setup \cite{expt}.  It is worth mentioning that multiplicities
were calculated keeping in mind the angular range covered by
combined Multics-Miniball array. Overall we can see that results
obtained with a \emph{soft} EoS are consistent with experimental
data at all colliding geometries. These findings reflects the
ability of molecular dynamics approaches (QMD, in our case) to
describe the reaction dynamics in low-energy regime.

Due to more explosive nature of hard EoS, spectator matter mainly
de-excites via emission of free nucleons and therefore, decline in
multiplicity of heavier clusters occurs. An increasing trend of
fragment multiplicity with centrality can be understood in terms
of more excitation energy deposited in spectator matter. In
semi-peripheral events, a larger chunk of excitation energy gets
transferred to spectator matter, thereby, leading to rise in
multiplicity of fragments with decrease in impact parameter.
Nuclear mean-field, therefore, becomes important factor governing
the outcome of spectator decay, while nucleon-nucleon collisions
dominate the participant matter physics. This analysis clearly
illustrates the relatively \emph{soft} nature of nuclear matter.

\section{\label{summary} Summary}
In summary, QMD model has been used to infer the inter-play of
different model inputs on fragment-emission in peripheral $^{197}$Au +$^{197}$Au collisions at 35 AMeV.
We find that choice of different nucleon-nucleon cross sections has marginal role to play at such a
low incident energy. However, the multiplicity of charged
particles obtained from the spectator decay are strongly
influenced by the incompressibility of the nuclear matter. The
hard equation of state results in enhanced emission of free
nucleons and fewer heavier fragments. Model calculations with soft
EoS are found to give encouraging results which are in accord with
experimental trends. This study favors \emph{soft} nature of the
nuclear matter. \\

\section{\label{ack} Acknowledgements}
One of the authors (Y. K. V) is acknowledges constructive
discussions with Drs. M. D' Agostino and M. Bruno. The research
grant from Indo-French Center for the Promotion of Advanced
Research (IFCPAR), New Delhi vide grant no. IFC/4104-1 is
gratefully acknowledged.

\newpage
\noindent {\Large \bf Figure Captions} \\

{\bf FIG. 1.} QMD simulation of Au (35 AMeV)+Au collisions at
reduced impact parameter $b/b_{max}$=0.55 (left panel) and
$b/b_{max}$=0.85 (right panel) as a function of time: (a) mean
nucleon density $\rho^{avg}/\rho_{o}$; (b) size of heaviest
fragment $A^{max}$; multiplicities of (c) free nucleons, (d) light
charged particles LCPs, and (e) fragments with mass $A\geq5$,
respectively. \\

{\bf FIG. 2.} Rapidity distribution dN/dy of free nucleons and
intermediate mass fragments (IMFS) as a function of scaled
transverse, $y^{(x)}/y_{beam}$ (left) and longitudinal,
$y^{(z)}/y_{beam}$ (right) rapidities in Au (35 AMeV)+Au reaction at
reduced impact parameter $b/b_{max}$=0.55. Solid and dashed curves
correspond to model calculations using a `soft' and a `hard' EoS
respectively. \\

{\bf FIG. 3.} The impact parameter dependence of multiplicity of
fragments with charge $3\leq Z \leq80$ obtained using a `soft' EoS
(solid line) and a `hard' EoS (dashed line) in Au(35 AMeV)+Au
collisions. Filled circles depict the experimental data points
\cite{expt}. \\

\newpage

\begin{figure}[!t]
\begin{center}
\includegraphics*[scale=0.70] {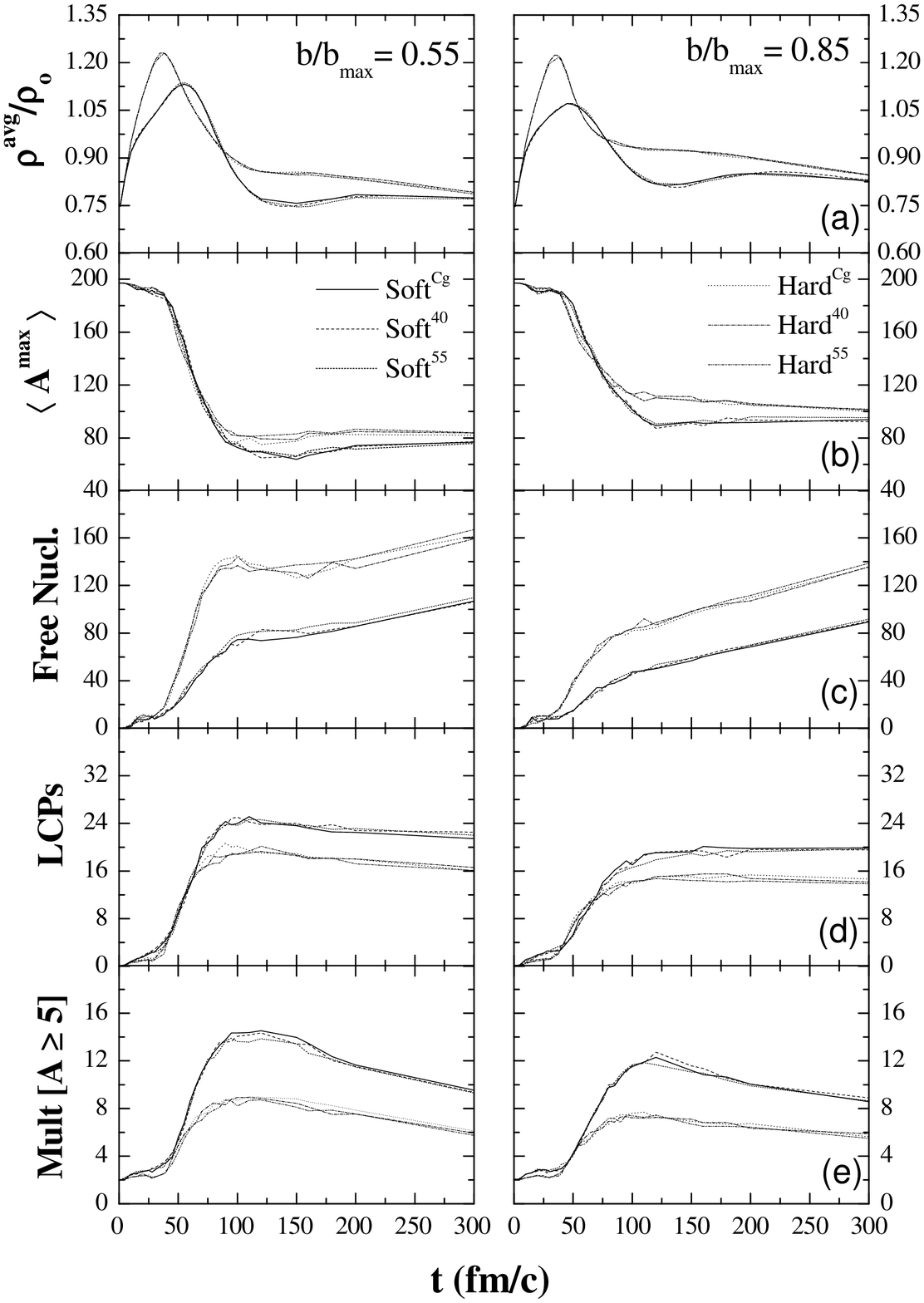} \caption{}
\end{center}
\end{figure}

\newpage

\begin{figure}[!t]
\begin{center}
\includegraphics*[scale=0.70] {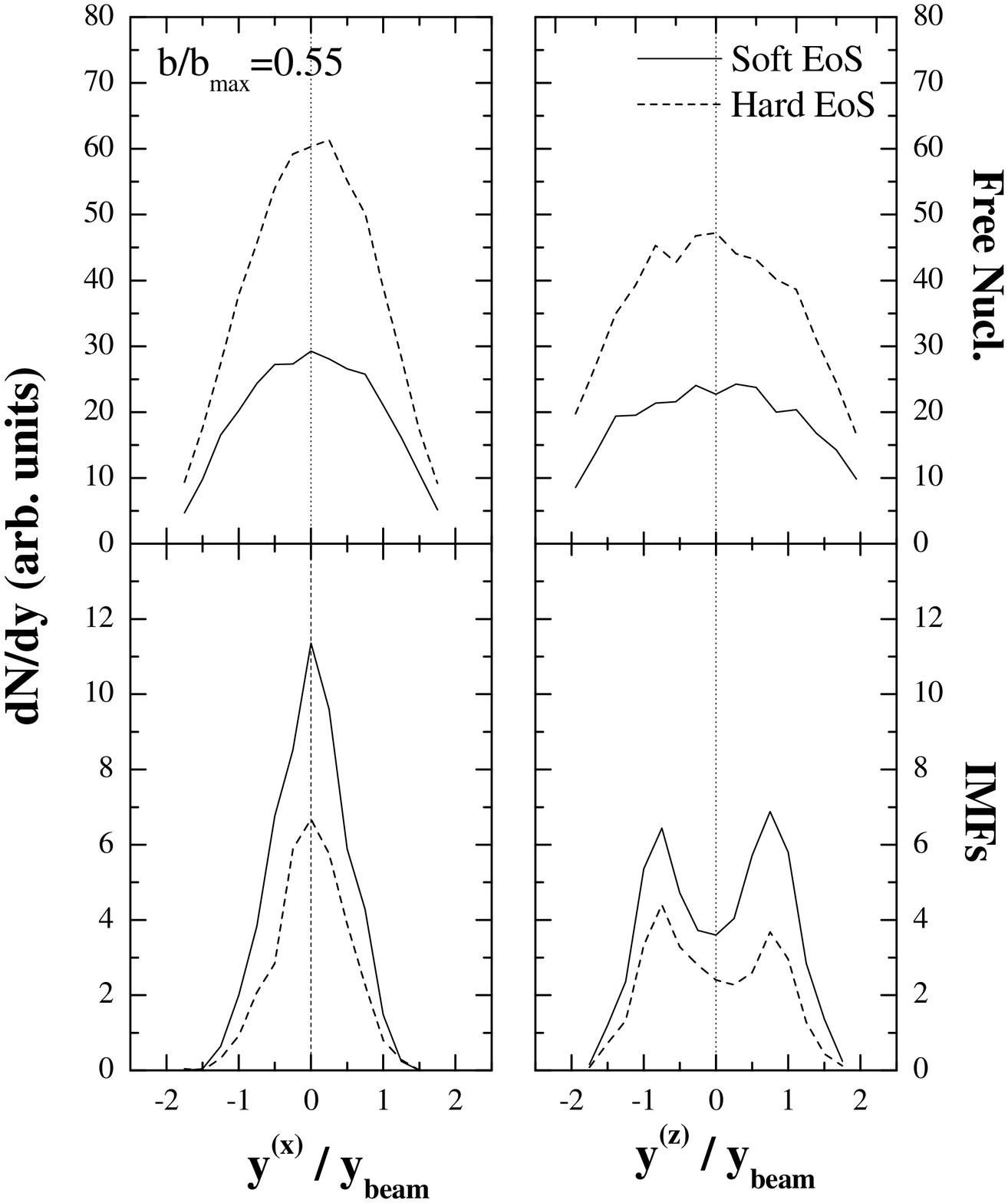} \caption{}
\end{center}
\end{figure}

\newpage

\begin{figure}[!t]
\begin{center}
\includegraphics*[scale=0.70] {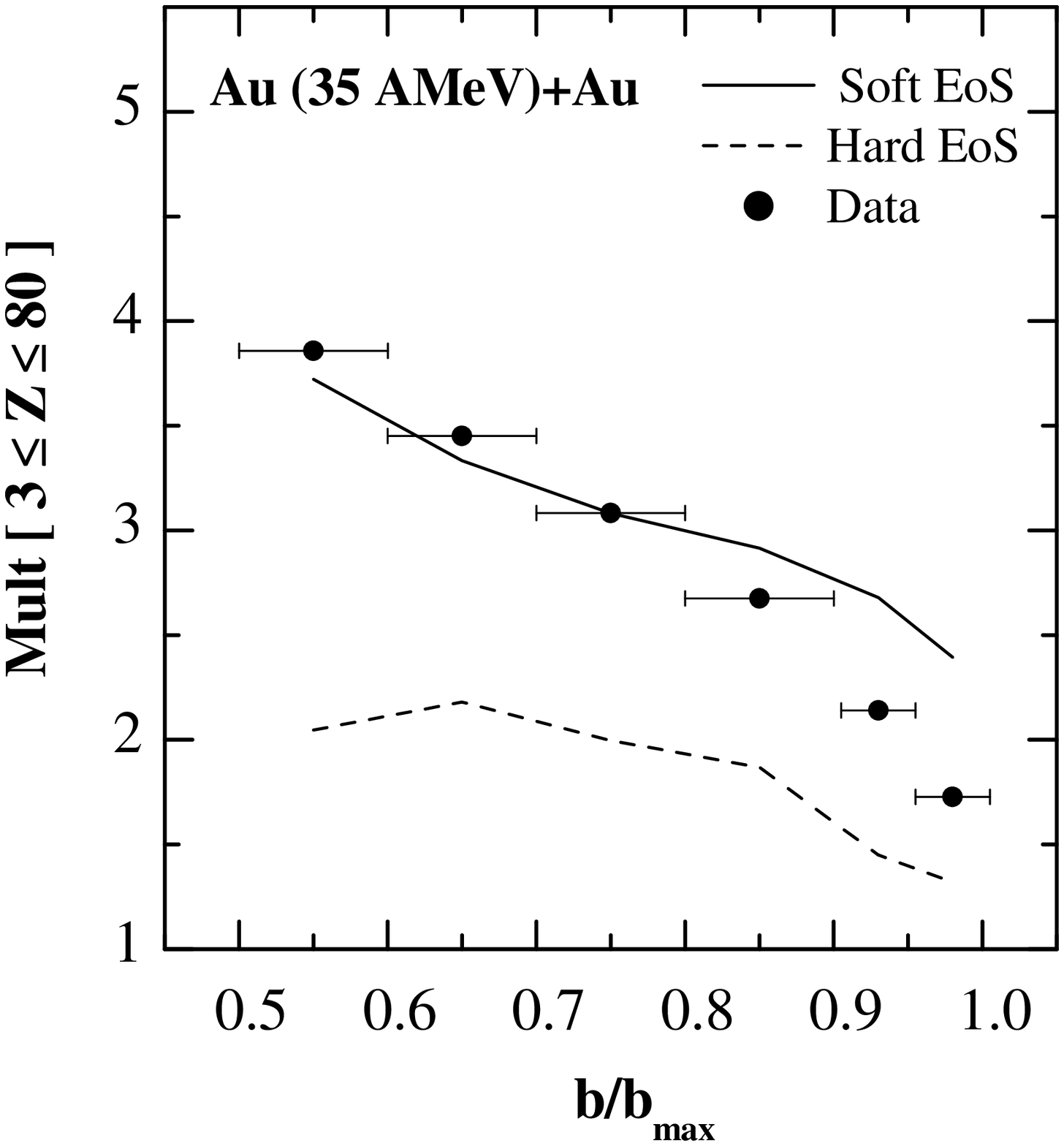} \caption{}
\end{center}
\end{figure}

\end{document}